\documentclass[aps,pra,preprint,groupedaddress,showpacs]{revtex4}
\usepackage{graphicx} 
\usepackage{dcolumn}  
\usepackage{bm}       
\begin{document}
\preprint{Version 3.0}

\title{Accurate S-state helium wave functions in momentum space}

\author{J. Sapirstein}
\email[]{jsapirst@nd.edu}

\affiliation{
Department of Physics, University of Notre Dame, Notre Dame, IN 46556}

\date{\today}

\begin{abstract}
High accuracy helium wave functions based on exponentials with
random coefficients are transformed into momentum space. The 
utility of the wave functions is demonstrated through calculation
of the expectation value of various operators needed to evaluate
relativistic and QED corrections.
\end{abstract}


\maketitle

\section{Introduction}

In recent years a basis set consisting of random exponentials has been used with 
increasing frequency to carry out calculations in helium \cite{Korobov}, \cite{Pachucki-S},
\cite{K-Yelkhovsky}. For singlet and triplet S-states it can be written as
\begin{equation}
\phi(\vec r_1, \vec r_2) =  \sum_i C_i 
[ e^{-\alpha_i r_1 -\beta_i r_2 - \gamma_i r_{12}}  \pm
e^{-\alpha_i r_2 -\beta_i r_1 - \gamma_i r_{12}} ],
\end{equation}
where $\alpha_i$, $\beta_i$, and $\gamma_i$ are parameters that are chosen
randomly in certain ranges, and the spin wavefunctions are understood.
Through careful choice of those ranges Korobov \cite{Korobov}
has been able to obtain a ground state energy of
\begin{equation}
E = -2.903~724~377~034~119~598~311~159~{\rm a.u.}.
\end{equation}
While even higher accuracies are possible with basis sets that incorporate
known nonanalytic behaviors of the wave function \cite{Schwartz}, the simple form
of the above wave function makes the evaluation of higher order corrections coming
from relativistic and QED corrections relatively straightforward. In this note we
will present calculations based on the momentum space form of Eq. 1, where the
wave function is defined through 
\begin{equation}
\phi( \vec p_1, \vec p_2) = \int {d^3 r_1 d^3 r_2 \over (2 \pi)^3} e^{-i \vec p_1 \cdot \vec r_1}
e^{-i \vec p_2 \cdot \vec r_2}  \phi(\vec r_1, \vec r_2).
\end{equation}
The Fourier transform is evaluated by first noting that
\begin{eqnarray}
e^{-\alpha r_1 -\beta r_2 - \gamma r_{12}}  & = & - {\partial^3 \over 
\partial \alpha \partial \beta \partial \gamma}
{e^{-\alpha r_1 -\beta r_2 - \gamma r_{12}} \over r_1 r_2 r_{12}} \nonumber \\
& = &  - {\partial^3 \over \partial 
\alpha \partial \beta \partial \gamma} \int {d^3 q_1 \over 2 \pi^2} 
{ e^{i \vec q_1 \cdot \vec r_1} \over \vec q_1\,^2 + \alpha^2}
\int {d^3 q \over 2 \pi^2} 
{ e^{i \vec q \cdot \vec r_{12}} \over \vec q^2 + \gamma^2}
\int {d^3 q_2 \over 2 \pi^2} 
{ e^{i \vec q_2 \cdot \vec r_2} \over \vec q_2\,^2 + \beta^2}.
\end{eqnarray}
To simplify the following discussion, we Fourier transform only the above expression,
and define it as $\phi_i(\vec p_1, \vec p_2)$, with the generalization to Eq. 1 being clear.
We see that
\begin{equation}
\phi_i(\vec p_1, \vec p_2) = - {1 \over \pi^3} {\partial^3 \over \partial 
\alpha \partial \beta \partial \gamma} 
\int d^3 q { 1 \over (\vec p_1 - \vec q)^2 + \alpha^2} {1 \over \vec q^2 + \gamma^2}
{ 1 \over (\vec p_2 + \vec q)^2 + \beta^2}.
\end{equation}
The integral over $d^3 q$ has been carried out analytically in Ref. \cite{Lewis}, and
leads to
\begin{equation}
\phi_i(\vec p_1, \vec p_2) = - {2 \over \pi} 
{\partial^3 \over \partial \alpha \partial \beta \partial \gamma} {\theta \over 
\sqrt{x}},
\end{equation}
where
\begin{eqnarray}
y & = & \gamma [ |\vec p_2 + \vec p_1|^2 + ( \alpha + \beta)^2] +
\beta(\gamma^2 + \alpha^2 + \vec p_1\,^2) + 
\alpha(\gamma^2 + \beta^2 + \vec p_2\,^2)  \nonumber \\
x & = & [ |\vec p_2 + \vec p_1|^2 + ( \alpha + \beta)^2] 
[ \vec p_1\,^2 + ( \alpha + \gamma)^2] 
[ \vec p_2\,^2 + ( \beta + \gamma)^2]  - y^2 \nonumber \\
\theta & = & {\rm arctan}({\sqrt{x} \over y}).
\end{eqnarray}
While this equation for $\phi_i ( \vec p_1 , \vec p_2)$ seems quite compact, the action of
the three derivatives leads to a considerably more complicated expression. It simplifies if we introduce
the auxiliary parameters
\begin{eqnarray}
X_1 & = & {d x \over d \alpha} ,  X_2 = {d x \over d \beta} , X_3 = { d x \over d \gamma} \nonumber \\  
Y_1 & = & {d y \over d \alpha} ,  Y_2 = {d y \over d \beta} , Y_3 = { d y \over d \gamma} \nonumber \\  
X_{23} & = & {d^2 x \over d \beta d \gamma} ,  X_{12} = {d^2 x \over d \alpha d \beta} , 
X_{13} = { d^2 x \over d \alpha d \gamma} \nonumber \\  
T_1 & = & X_1 X_{23} + X_2 X_{13} + X_3 X_{12}, T_2 = X_1 + X_2 + X_3,
T_3 = Y_1 X_{23} + Y_2 X_{13} + Y_3 X_{12} \nonumber \\
T_4 & = & Y_1 + Y_2 + Y_3, T_5 = X_1 Y_2 Y_3 + X_2 Y_1 Y_3 + X_3 Y_1 Y_2 \nonumber \\
T_6 & = & Y_1 X_2 X_3 + Y_2 X_1 X_3 + Y_3 X_1 X_2.
\end{eqnarray}
In terms of these we find
\begin{eqnarray}
\phi_i (\vec p_1, \vec p_2) & = & -{2 \over \pi} [ {3 \theta \over 8  x^{7/2}} [ 2 x T_1 - 5 X_1 X_2 X_3]
+ { y'' (T_2 + 2 y T_4) \over D^2} + {T_3 \over D^2} \nonumber \\
& - & { 4 y T_5 \over D^3} + {2 (x-3y^2)Y_1Y_2Y_3 \over D^3} +
{ (40xy^3 + 33 x^2 y + 15 y^5)X_1 X_2 X_3 \over 8 x^3 D^3} \nonumber \\
& - & { y T_1(3y^2 + 5x) \over 4 x^2 D^2} - { 2 T_1 \over D^3} - {2 \over D}],
\end{eqnarray}
where $D = x+ y^2$ and $y'' = 2(\alpha + \beta + \gamma)$. We note the following symmetry properties
of the momentum space wave function. Because the basic functions that form the wave
function, $x$ and $y$, are invariant under the simultaneous replacement $\vec p_1 \leftrightarrow
\vec p_2$ and $\alpha \leftrightarrow \beta$, for singlet states $\phi(\vec p_1, \vec p_2) =
\phi(\vec p_2, \vec p_1)$ and for triplet states $\phi(\vec p_1, \vec p_2) = -\phi(\vec p_2, \vec p_1)$. 
Both symmetries also hold when
the magnitudes of $\vec p_1$ and $\vec p_2$ are switched.

There are various uses for the wave function of helium in momentum space, notably
application to scattering calculations. While most work has been carried out with
Fourier transforms of Hartree-Fock wave functions, an approach that includes 
correlation more completely is given in Ref. \cite{scatref}. In that work, a Hylleraas
basis set was Fourier transformed and applied to the calculation of a
number of helium properties, notably the Compton profile. While the wave function used
here gives a more accurate energy (odd powers of $r_{12}$ were not included in the
basis set of Ref. \cite{scatref}), we do not find appreciably different answers for 
any of the quantities calculated there. In particular, we find the same small difference
between using fully correlated wave functions and Hartree-Fock wave functions for the
Compton profile. Rather than pursuing this line of research we instead now
discuss applications of the momentum space wave function to the calculation of
higher order relativistic and QED corrections to the energies of the $1^1S_0$, $2^1S_0$,
and $2^3S_1$ states of neutral helium.

These corrections are obtained from a set of operators $O_i$, with associated energy
shifts $E_i = \langle \phi|O_i|\phi\rangle$. The first set of operators was derived by Breit \cite{Breit1},
and describe corrections of order $\alpha^2$ a.u.. The equation he used in the derivation has
certain difficulties connected with negative energy states \cite{Brown}, but later treatments
using the Bethe-Salpeter equation \cite{Araki}, \cite{Sucher}, \cite{DK} treat
negative energy states consistently and allow the systematic treatment of higher order
corrections.

Most recent calculations, however, use the technique of effective field theory \cite{Caswell-Lepage}
to derive the operators. We note in particular the compact rederivation of
the Douglas-Kroll \cite{DK} results for contributions to the fine structure of helium in order
$\alpha^4$ a.u. of Ref. \cite{Pach1}, and the derivation of contributions to the energy of the
ground state to the same order in Ref. \cite{Yelkhovsky}.
The idea of effective field theory is to compare free-particle scattering amplitudes in QED to
an effective nonrelativistic theory, with operators added to the Schr\"{o}dinger Hamiltonian
that account perturbatively for the difference of the amplitudes. Once the operators have
been determined from considering free-particle scattering, they are used as perturbations
in standard bound-state Rayleigh-Schr\"{o}dinger perturbation theory to calculate energy shifts.
In this approach it is
natural to work in momentum space, and the operators are then Fourier transformed to coordinate
space for numerical evaluation. Here, however, because we have formed wave functions in momentum
space, we avoid this step, and work exclusively in momentum space. This has the advantage of
simplicity, but the disadvantage of being less accurate than coordinate space. We now give
a brief rederivation of the Breit operators using effective field theory,
and  illustrate their numerical evaluation in momentum space.

The connection of the scattering amplitude of two electrons with momentum $p_1$, $p_2$ to
scatter into states with momentum $p_3$, $p_4$ to an energy shift is
\begin{equation}
E_i = {1 \over (2 \pi)^{6}} \int d^3 p_1 d^3 p_2 d^3 p_3 d^3 p_4
\phi^*(\vec p_3, \vec p_4) O_i(\vec p_3, \vec p_4; \vec p_1, \vec p_2) \phi(\vec p_1, \vec p_2).
\label{ebasic}
\end{equation}
We work in the center of mass frame, with initial nuclear momentum $-\vec p_1 - \vec p_2$,
final nuclear momentum $-\vec p_3 - \vec p_4$. If only electron-electron interactions (ee) 
are considered, momentum conservation allows us to write
\begin{equation}
{O_i}^{ee}(\vec p_3, \vec p_4; \vec p_1, \vec p_2) = (2 \pi)^3 \delta^3 (\vec p_3 + \vec p_4 -
\vec p_1 - \vec p_2) {M_i}^{ee}(\vec p_3; \vec p_1, \vec p_2)
\end{equation}
with the associated energy shift
\begin{equation}
{E_i}^{ee} = {1 \over (2 \pi)^{3}} \int d^3 p_1 d^3 p_2 d^3 p_3 
\phi^*(\vec p_3, \vec p_1 + \vec p_2 - \vec p_3) {M_i}^{ee}(\vec p_3; \vec p_1, \vec p_2) 
\phi(\vec p_1, \vec p_2).
\label{ebasicex}
\end{equation}
If instead we consider diagrams in which one of the electrons, taken to be electron 1,
interacts with the nucleus (eN), and electron 2 does not participate, we can write
\begin{equation}
{O_i}^{eN}(\vec p_3, \vec p_4; \vec p_1, \vec p_2) = (2 \pi)^3 \delta^3 (\vec p_4 -
\vec p_2) {M_i}^{eN}(\vec p_3, \vec p_1)
\end{equation}
with the energy shift
\begin{equation}
{E_i}^{eN} = {1 \over (2 \pi)^{3}} \int d^3 p_1 d^3 p_2 d^3 p_3 
\phi^*(\vec p_3, \vec p_2 ) {M_i}^{eN}(\vec p_3, \vec p_1) 
\phi(\vec p_1, \vec p_2).
\label{ebasicnex}
\end{equation}
Diagrams with all three particles interacting have no delta functions, and have
to be evaluated with the 12-dimensional integral of Eqn. \ref{ebasic}. In all cases
we note that 3 of the integration variables can be carried out trivially, and that
the adaptive Monte-Carlo program VEGAS \cite{VEGAS} can be used to numerically
evaluate the integrals, though with far less accuracy than available from coordinate
space techniques. This loss of accuracy is due entirely to the fact that the multidimensional
integrals have to be carried out numerically: the wave functions themselves are quite
accurate. In the calculations presented here we use 600 basis functions, and the energy
eigenvalues are accurate to more than 14 digits for the ground state and 11 for the excited
S states.

We begin by evaluating the relativistic mass increase (RMI) operator, which we
treat as an ee diagram. The contribution to the scattering amplitude of this
operator is
\begin{equation}
{M_{RMI}}^{ee}(\vec p_3; \vec p_1, \vec p_2) = - (2 \pi)^3 {{p_1}^4 + 
{p_2}^4 \over 8 m^3} \delta( \vec p_3 - \vec p_1),
\end{equation}
which gives the energy shift
\begin{equation}
{E_{RMI}}^{ee} = - { 1 \over 8 m^3}  \int d^3 p_1 d^3 p_2 
\phi^*(\vec p_1, \vec p_2) (p_1^4 + p_2^4) \phi(\vec p_1, \vec p_2).
\end{equation}
The result is tabulated in the first row of Table 1. 

We next turn to corrections to Coulomb scattering between
the electrons. In this case the nonrelativistic scattering
operator is
\begin{equation}
{M_{C}}^{ee}(\vec p_3; \vec p_1, \vec p_2) = { 4 \pi \alpha \over |\vec p_3 - \vec p_1|^2}.
\end{equation}
This corresponds to the coordinate space potential $\alpha / | \vec r_1 - \vec r_2|$. 
To calculate relativistic corrections, we use Dirac spinors to describe scattering and work
in Coulomb gauge. We introduce the notation DC to refer to the scattering amplitude with
exchange of a Coulomb photon, and DT for the scattering
amplitude with exchange of a transverse photon.
The DC scattering amplitude can then be Taylor expanded in powers of $p/m$, with the leading
corrections given by
\begin{eqnarray}
{M_{\rm DC}}^{ee}(\vec p_3; \vec p_1, \vec p_2) = {M_C}^{ee}( \vec p_3; \vec p_1,\vec p_2)
 [ 1 -{ | \vec p_3 - \vec p_1|^2 \over 8 m^2} 
-{ | \vec p_4 - \vec p_2|^2 \over 8 m^2} + \nonumber \\
{ i \vec \sigma_1 \cdot (\vec p_3 \times 
\vec p_1) \over 4 m^2} + { i \vec \sigma_2 \cdot (\vec p_4 \times \vec p_2) \over 4 m^2} ],
\label{eeexch}
\end{eqnarray}
with the understanding that $\vec p_4 = \vec p_1 + \vec p_2 - \vec p_3$.
The first two correction terms are Darwin terms, and sum to $\pi \alpha/m^2 \delta^3(\vec r_2 -
\vec r_1)$ in coordinate space. The last two are spin-orbit operators, which do not contribute to the
S-states considered here. We tabulate the Darwin terms in the second row of Table 1.

Considering now Coulomb scattering between an electron and the nucleus, which we take
to have charge $Z$ although only $Z=2$ will be considered here,
the nonrelativistic limit is given by
\begin{equation}
{M_{C}}^{eN}(\vec p_3, \vec p_1) = -{ 4 \pi \alpha Z \over |\vec p_3 - \vec p_1|^2} 
\end{equation}
with associated energy shift
\begin{equation}
{E_C}^{eN} = -{1 \over (2 \pi)^{3}} \int d^3 p_1 d^3 p_2 d^3 p_3 \phi^*(\vec p_3, \vec p_2 ) 
{ 4 \pi \alpha Z \over |\vec p_3 - \vec p_1|^2} \phi(\vec p_1, \vec p_2).
\end{equation}
Relativistic corrections are now obtained by introducing a Dirac spinor only for the electron
(the nucleus is treated here in the infinite mass limit), and we find for exchange of a Coulomb
photon
\begin{equation}
{M_{\rm DC}}^{eN}(\vec p_3, \vec p_1) = {M_C}^{eN}(\vec p_3, \vec p_1) 
[ 1 -{ | \vec p_3 - \vec p_1|^2 \over 8 m^2} + { i \vec \sigma_1 \cdot (\vec p_3 \times 
\vec p_1) \over 4 m^2}  ]
\end{equation}
Again only the Darwin term contribute for S-states, but now corresponds to $ \pi Z \alpha /m^2
\delta^3(\vec r_1)$,
which we tabulate in the third row of Table 1.

The effect of transverse photon exchange between the electrons is simplified if we neglect retardation,
which enters in order $\alpha^3$ a.u., and in this approximation we have
\begin{eqnarray}
{M_{\rm DT}}^{ee}(\vec p_3; \vec p_1, \vec p_2) = {\pi \alpha \over q^4} - { \pi \alpha \over m^2 q^2}(\vec p_1 + \vec p_3)
\cdot (\vec p_2 + \vec p_4) - {\pi \alpha \over m^2}(\vec \sigma_1 \cdot \vec \sigma_2 -
\vec \sigma_1 \cdot \hat{q} \vec \sigma_2 \cdot \hat{q}) \nonumber \\
- { 2 \pi i \alpha \over m^2 q^2} \vec \sigma_1 \cdot ( \vec p_2 \times \vec p_4) +
{ 2 \pi i \alpha \over m^2 q^2} \vec \sigma_2 \cdot ( \vec p_3 \times \vec p_1) ,
\end{eqnarray}
where $q = | \vec p_3 - \vec p_1|$.
The first two terms, referred to as orbit-orbit terms, or as $E^{(2)}$,
are usually evaluated by Fourier transforming into coordinate space. With the present
approach, however, they are quite easily treated, and the result presented in the
fourth row of Table 1. The third term is another delta function, and the last terms again 
vanish for the S-state considered here. The effect of transverse photon exchange for eN 
scattering vanishes in the infinite nuclear mass limit used in this work.

The operators considered so far in this note have been studied for many decades, and have
all been evaluated with much higher accuracy than presented here \cite{Drake0}. The utility of the
present approach lies in the fact that operators that enter in higher order, generally
derived in momentum space, are both fairly complicated when Fourier transformed to coordinate 
space, and in addition need to be evaluated with less accuracy than the terms treated above. We 
illustrate this point with relativistic corrections that contribute in order $m \alpha (Z \alpha)^6$ to
S-states. A complete set of operators for triplet states has been derived by Pachucki \cite{Pach2s} using
an effective field theory approach, and we consider here the corrections to one-Coulomb photon
exchange, Eq. 20 of that paper,
\begin{eqnarray}
V_1 = {4 \pi \alpha \over q^2} {1 \over 64 m^4} [ q^4 - {4 \over 3} ( \vec p_3 \times
\vec p_1)\cdot(\vec p_4 \times \vec p_2) (\vec \sigma_1 \cdot \vec \sigma_2) + \nonumber \\
{5 \over 2} [ (p_4^2 - p_2^2)^2 + ( p_3^2 - p_1^2)^2] + 3 q^2(p_1^2 + p_3^2 + p_2^2 + p_4^2)].
\end{eqnarray}
This expression corresponds to the next term in the $p/m$ expansion in Eq. \ref{eeexch}.
While Ref. \cite{Pach2s} treats triplet states, this particular result is also valid for the
singlet case. We note that the last term vanishes for triplet states because of the 
symmetry $\phi(\vec p_1, \vec p_2) = - \phi(\vec p_2, \vec p_1)$ mentioned earlier, as also noted in 
Ref. \cite{Pach2s}. The resulting energy shift of the $2^3S_1$ state is
\begin{equation}
E^{ee} = -0.006 2(2) m \alpha^6.
\end{equation}
While again not of high accuracy, we note the extremely simple nature of the coding, which
is almost identical to the program that evaluates the Darwin term.
This contrasts with the more complicated coordinate
space calculation, where numerous derivatives must be applied to wave functions, leading to
a much lengthier expression. As with the $m \alpha^4$ corrections, much higher
accuracy is available from working in coordinate space, with -0.006 344 7 $m \alpha^6$
the known result \cite{private}. However, we note the momentum space accuracy corresponds to
3.7 kHz, to be compared with the experimental accuracy \cite{Shiner1} of 60 kHz.

While the formula for $V_1$ given above is valid for singlet states, it gives a linearly divergent
result in that case. It is quite simple, however, to regulate this 
divergence in momentum space, where one simply imposes the cutoff $|\vec p_i| < \Lambda$. An 
application of this momentum space regulator
to the case of ground state positronium hyperfine splitting can be found in \cite{Zeb}. In Table II we
show results for the expectation value of $V_1$ for the ground state of helium with different
cutoffs $\Lambda$, with the linear dependence on $\Lambda$ clearly visible.
When combined with other linearly divergent terms in a complete calculation
a $\Lambda$ independent result will obtain in the limit of large $\Lambda$. By
improving the accuracy found in this part of the calculation 
this procedure can be used to check the results of Ref. \cite{K-Yelkhovsky} 
without explicitly canceling the divergences: work on this problem is in progress.

\section{Conclusions}

We have presented the formula for the momentum space form of a powerful basis set for helium.
While it has the potential for proving useful for scattering calculations on helium, we have
concentrated on evaluating expectation values of operators that give relativistic and QED
corrections to energy levels. Because these operators are derived in momentum space, this
allows one to work entirely in momentum space. The next step in this research is the extension
to states with nonvanishing angular momentum. The most important application we have in mind
is to the fine structure of helium P states, where recent high-precision measurements by Hessels and
collaborators \cite{Hessels} have the potential of allowing a determination of the fine
structure constant $\alpha$ to a precision of 4 ppb. Unfortunately, the present state of theory
is unclear, where the most complete calculation by Drake \cite{Drake}, while consistent with the 
fine structure interval $\nu_{01}$ measured in Ref. \cite{Hessels}, is
inconsistent with measurements of the interval $\nu_{12}$ \cite{Hessels1}, \cite{Shiner2}.
This inconsistency has also been noted in Ref. \cite{Pachucki-S}.
It is possible that the relative simplicity of the method developed here can shed light on this 
situation.

\begin{acknowledgments}
This work was supported in part by NSF Grant No.~PHY-0097641. Conversations
with K. Pachucki and S. Morrison are gratefully acknowledged, with particular
thanks to the former for providing details of his $2^3S_1$ calculation and
helpful comments on the manuscript.
\end{acknowledgments}

\newpage

\begin{table}
\caption{Expectation of operators for n=1 and 2 S states of helium in units of $\alpha^2$ a.u..
The notation $H_2$ in ${E_{DT}}^{ee}$ indicates the delta function is not included
in the result.}
\begin{tabular}{lrrr}
Operator  &  $1^1S_0$~~~   & $2^1S_0$~~~    &    $2^3S_1$~~~     \\
\hline
${E_{RMI}}^{ee}$             & -13.5212(3)~~    &  -10.27959(5)~~   &  -10.45887(4)       \\
${E_{DC}}^{ee}$        &  0.3346(3)~~    &  0.02718(8)~~   &  0.0       \\
${E_{DC}}^{eN}$        &  5.6879(2)~~    &  4.1139(2)~~   &  4.1479(2)      \\
${E_{DT}}^{ee}(H_2)$  & -0.1393(2)~~     & -0.00922(1)~~   & -0.00157(7)       \\
\end{tabular}
\end{table}

\begin{table}
\caption{Expectation value of $V_1$ for the ground state of helium with the 
regulator $|\vec p_i| < \Lambda$ for different values of $\Lambda$. Units
$\alpha^4$ a.u. for $\langle V_1 \rangle$ and $m\alpha$ for $\Lambda$.}
\begin{tabular}{lrrr}
$\Lambda$ &  100~~~~~            & 200~~~~~               &    300~~~     \\
\hline
$\langle V_1 \rangle$   & -26.9(1)~~   &  -59.4(4)~~   &  -90.9(6)       \\
\end{tabular}
\end{table}

\end{document}